# Magnetic and dielectric behavior in $YMn_{1-x}Fe_xO_3$ ($x \leq 0.5$)

Neetika Sharma, A. Das*, S.K. Mishra, C.L. Prajapat+, M.R. Singh+, S.S. Meena

*Solid State Physics Division, Bhabha Atomic Research Centre, Mumbai 400085, India*

+*Technical Physics Division, Bhabha Atomic Research Centre, Mumbai 400085, India*

**Abstract**

The role of doping Fe on the structural, magnetic and dielectric properties of frustrated antiferromagnet $YMn_{1-x}Fe_xO_3$ ($x \leq 0.5$) has been investigated. The neutron diffraction analysis shows that the structure of these polycrystalline samples changes from hexagonal phase (space group $P6_3cm$) to orthorhombic phase (space group *Pnma*) for x > 0.2. The frustration parameter decreases with Fe substitution. All the compounds are antiferromagnetic and the magnetic structure is described as a mixture of Γ3 and Γ4 irreducible representation (IR) in the hexagonal phase and the ratio of these two IRs is found to vary with Fe doping (x ≤0.2). A continuous spin reorientation as a function of temperature is observed in these samples. The magnetic ground state in the orthorhombic phase of the higher doped samples (x ≥0.3) is explained by taking Γ1 ($G_xC_yA_z$) representation of *Pnma* setting. In $YMnO_3$ suppression of dielectric constant ε´ is observed below $T_N$ indicative of magnetoelectric coupling. This anomalous behavior reduces in Fe doped samples. The dielectric constant is found to be correlated with the magnetic moment (M) obtained from neutron diffraction experiments and follows a $M^2$ behavior close to $T_N$ in agreement with Landau theory.

## Introduction

Hexagonal manganites are fascinating materials displaying ferroelectricity and magnetism in a single phase. These compounds are paraelectric (space group $P6_3/mmc$) at high temperature but they undergo a structural transition to ferroelectric phase (space group $P6_3cm$) below T ~ 1000 K [1]. The hexagonal manganites undergo ferroelectric transition far above the magnetic ordering of $Mn^{3+}$ ions. Buckling of $MnO_5$ trigonal bypiramids and the displacement of the $R^{3+}$ ions are the two factors which have been shown to be responsible for the ferroelectric polarization [2]. Amongst this, $YMnO_3$ is one of the most studied geometric frustrated hexagonal mangnaites. Due to the large difference between the antiferromagnetic transition temperature ($T_N$ ~ 75 K) and ferroelectric transition temperature ($T_{FE}$ ~ 950 K) this compound is classified as Type I multiferroic compound [3] indicative of a weak coupling between the two ordering. The in-plane dielectric constant though shows a distinct anomaly at $T_N$ [4-6] thereby, indicating coupling between the electric polarization and magnetic ordering. A large and sharp decrease of the dielectric constant below $T_N$ has been observed and is attributed to charge-transfer excitation in the geometrical frustrated system [7]. Each Mn atom in these compounds is surrounded by three in-plane and two apical oxygen atoms, thus forming the $MnO_5$ trigonal bypiramid and the $Mn^{3+}$ ions also form natural 2D network of corner sharing triangular network which leads to frustration. A measure of such frustration is defined as a ratio of $\frac{|\theta_{CW}|}{T_N}$ (where $\theta_{CW}$ is the Curie-Weiss temperature) and is ~ 6 for this compound [8]. The magnetic structure of $YMnO_3$ could be explained by considering either Γ1 or Γ3 irreducible representations (IR) [9, 10]. But according to recent experimental studies the magnetic structure of $YMnO_3$ is best explained by linear combination of $\Gamma_3 + \Gamma_4$ IR of space group $P6_3cm$ [11, 12].

By doping at rare earth site or with the application of external pressure one can continuously change the magnetic ground state [13-17]. Previous studies have shown that the substitution of Mn by non-magnetic ions $Ti^{4+}$ and $Ga^{3+}$, changes the ferroelectric properties [18,19] and large enhancement in magnetocapacitance has been found in both Ti and Ga doped samples. The role of possible changes in magnetic structure leading to large magnetocapaciatance has been raised. In another study it has been shown that the value of ferroelectric polarization can be controlled by changing the magnetic ground state [20]. In our previous study we have shown that the magnetic structure of the Ga doped compound remains same albeit with a reduced value of ordered moment while Ti doping at Mn site changes the magnetic structure to $\Gamma_2$ IR [12]. Large magnetodielectric effects have also been observed coinciding with the spin reorientation temperature in $HoMnO_3$ [21]. In our previous investigation we had reported that Fe doping at Mn site leads to observation of reorientation of Mn magnetic moments as a function of temperature. The modified magnetic structure can be described as a linear combination of irreducible representation $\Gamma3+\Gamma4$ with different mixing ratio of these two representations [12] raising the possibility of observation of dielectric anomalies at the spin reorientation temperatures.

The substitution of Fe by Mn or Mn by Fe is interesting because of the same ionic radii of $Fe^{3+}$ and $Mn^{3+}$ yet different magnetic moment values. Because of the presence of an electron in $dz^2$ orbital in case of $Fe^{3+}$, substitution of $Fe^{3+}$ allow us to study the effect of electron doping in the $YMnO_3$. In our earlier studies of Fe doping at Mn site, we have observed reduction in $T_N$ as well as spin reorientation behavior as a function of temperature [12]. This behavior is opposite of that which has been reported for $YbMn_{1-x}Fe_xO_3$ [22, 23]. With increase in Fe concentration in $YbMnO_3$, enhancement in $T_N$ has been observed. The neutron diffraction data reported in this work were not analyzed for the resulting change in magnetic structure, if any. In the opposite end, in orthorhombic $YFe_{1-x}Mn_xO_3$ ($0.10 \leq x \leq 0.45$) it has been found that the magnetic structure changes from $\Gamma_4$ ($A_x F_y G_z$) at high temperature to $\Gamma_1$ ($G_x C_y A_z$) at low temperature, using magnetic torque measurements [24].The spin reorientation in Mn substituted orthoferrites has been attributed to the large magnetic anisotropy energy of the $Mn^{3+}$ ion at low temperature. The direction of antiferromagnetic spin axis in $YFeO_3$ is due to the anisotropy energy of the $Fe^{3+}$ ions, but the substitution of Mn at Fe site after the critical concentration overcome the anisotropy energy of the $Fe^{3+}$ ions. This spin reorientation behavior in orthorhombic $YFe_{1-x}Mn_xO_3$ has been recently confirmed by neutron diffraction studies [25]. Dielectric anomaly near the magnetic transition temperature appeared in orthorhombic $YFe_{1-x}Mn_xO_3$ and it has been observed that the dielectric anomaly is more pronounced at higher Mn concentration. The origin of magnetodielectric coupling in these compounds is attributed to the spin-phonon coupling [26]. However, there is an absence of similar work in the hexagonal rich end of the series, $YMn_{1-x}Fe_xO_3$. The few studies that has been carried out on Fe doping at Mn site, report the concentration range over which the structure changes to orthorhombic [27, 28]. In the present work we have synthesized $YMn_{1-x}Fe_xO_3$ ($0.1 \leq x \leq 0.5$) and report the effects of $Fe^{3+}$ ($d^5$) doping on the structural, dielectric and magnetic properties of frustrated hexagonal $YMnO_3$. We find that with Fe doping at Mn site progressive changes in chemical structure and magnetic structure as a function of temperature and composition. Large dielectric anomalies are observed coinciding with the magnetic transition temperature in these compounds which we show to scale with sub lattice magnetization obtained from neutron diffraction, in agreement with predictions of Landau theory.

## Experimental Details

Polycrystalline samples of $YMn_{1-x}Fe_xO_3$ (x = 0.1 - 0.5) were synthesized by conventional solid state reaction route. The starting materials $Y_2O_3$, $MnO_2$, and $Fe_2O_3$ were mixed in stoichiometric ratio and calcined in air at 1200°C for 90 hrs with several intermediate grindings. The phase purity of these samples was confirmed by x-ray powder diffraction recorded on a Rigaku diffractometer, using Cu Kα radiation in the angular range $10^o \leq 2\theta \leq 70^o$ at room temperature. The magnetization measurements were carried out on a Superconducting quantum interference design magnetometer (SQUID). The zero field and field cooled measurements were performed under a magnetic field of 0.1 T. For dielectric measurements, disk-shaped pellets of approximately16 mm in diameter and 1–2 mm in thickness were prepared using a uniaxial isostatic press. Silver paste was applied on the polished surfaces of the disks to form the electrodes. Low-temperature dielectric measurements were carried out using a frequency-response analyzer (Novocontrol *TB*-Analyzer). For cooling the sample down to 5 K, a closed-cycle refrigerator with He-gas exchange attachment was used. Temperature-dependent capacitance data was measured in the frequency range 1Hz–1MHz with a heating rate of 0.8K/min in the range from 5 to 300K. The neutron diffraction patterns were recorded on a multi-PSD-based powder diffractometer ( $\lambda$ = 1.2443Å ) at the Dhruva reactor, Bhabha Atomic Research Centre, Mumbai, at selected temperatures between 6K and 300K in the angular range $5^o \leq 2\theta \leq 140^o$. The patterns were refined by the Rietveld refinement technique using FULLPROF program [29].

## Results and Discussion

The parent compound ($YMnO_3$) and the end compound ($YFeO_3$) of this series have different structures. $YMnO_3$ crystallizes in hexagonal phase with space group $P6_3cm$ whereas $YFeO_3$ crystallizes in an orthorhombically distorted perovskite structure with space group *Pnma*. In hexagonal phase (space group $P6_3cm$) $Mn^{3+}$ ions are in five-fold coordination, surrounded by oxygen ions forming trigonal bypiramids and $Y^{3+}$ ions are in seven-fold coordination. The $MnO_5$ trigonal bypiramids are two dimensionally arranged in space and are separated by a layer of $Y^{3+}$ ions. In perovskites having orthorhombic phase (space group *Pnma*), $Mn^{3+}$ ions are in six-fold coordination in the center of octahedral and rare earth ions (R) ions are nine coordinated. All the samples have been characterized by x-ray and neutron diffraction techniques. From Rietveld refinement of these diffraction patterns a single hexagonal phase has been observed for $YMn_{1-x}Fe_xO_3$ for $x \leq 0.2$. For $x > 0.2$ the diffraction lines belonging to the orthorhombic phase (space group *Pnma*) characteristic of $YFeO_3$ compound, appear in the x-ray diffraction pattern and therefore analysis with a mixed orthorhombic and hexagonal phase has been carried out for these samples ($0.3 \leq x \leq 0.5$). With increase in Fe doping, there is a progressive increase of the orthorhombic phase as shown in figure 1. This is in agreement with previous studies where structural transition in $YMnO_3$ has been observed by doping at Y or Mn site [13, 16, 30-31] The variation of cell parameters and volume for a single phase hexagonal sample $YMn_{0.8}Fe_{0.2}O_3$ is shown in figure 2(a) and 2(b) respectively. For hexagonal YMnO3, it has been observed that the lattice parameter a increases and c decreases with increase in temperature and similar behavior of parameters has been observed in all the studied samples in hexagonal phase. The negative thermal expansion in of c parameter in hexagonal manganites is explained by the reduction of tilting of $MnO_5$ bypiramids along with the buckling of Y-planes [1]. The unit cell volume decreases with temperature. However, they exhibit an anomalous behavior in the magnetically ordered state which might be attributed to the magnetoelastic coupling in these hexagonal compounds. The tilting and buckling of $MnO_5$ polyhedra are important lattice distortion parameters for the hexagonal $YMnO_3$. The tilting of $MnO_5$ bypiramid is represented by the angle (α) between the O1-O2 axis and c axis and the buckling is represented by the angle β between the O3-O4-O4 plane and c axis [32]. Doping at Y or Mn site is expected to modify the distortion parameters of $MnO_5$ bypiramid [33]. We find that these distortion parameters are suppressed with Fe doping at Mn- site and are shown in inset of figure 2(a) and 2(b). This reduction of tilting and buckling in Fe doped samples indicates the suppression of average interaction [30] within Mn trimers and this is evidenced in decreased magnetodielctric coupling in Fe doped sample (shown below). Figure 3, shows the temperature variation of unit cell volume of the orthorhombic phase for $YMn_{0.6}Fe_{0.4}O_3$. The unit cell volume increases continuously with increase in temperature. The temperature dependence of volume has been fitted to Debye model, using Grüneisen approximation [34]. In the Grüneisen approximation, the temperature dependence of volume is described by, $V(T) = \gamma\, U(T)/B_0 + V_0$, where $\gamma$, $B_0$, and $V_0$ are the Grüneisen parameter, bulk modulus and volume, respectively, at T = 0 K. In the Debye model, internal energy U (T) is given by, $U(T) = 9Nk_BT\left(\frac{T}{\theta_D}\right)^3 \int_0^{\frac{\theta_D}{T}} \frac{x^3}{e^x - 1}dx$ , where N is the number of atoms in the unit cell, $k_B$ is the Boltzmann's constant and $\theta_D$ is the Debye temperature. The red curve in the figure 3 represents the volume obtained by fitting the volume data to the Debye- Grüneisen equation. It is seen from the fit, that the temperature dependence of the volume does not show any anomalous behavior. The refined parameters obtained by neutron diffraction at 6 K are given in Table 1.With Fe doping, change in Mn-O bond lengths has been observed in hexagonal phase. The $Mn-O_1$ bond lengths decrease with Fe concentration while enhancement in $Mn-O_2$ and $Mn-O_3$ bond lengths is observed.

Figure 4 shows the temperature dependence of magnetization, M(T) for $YMn_{1-x}Fe_xO_3$ (x = 0.1-0.5) under an applied magnetic field of 0.1 T. It has been observed that the parent sample $YMnO_3$ does not exhibit a distinct anomaly at the transition temperature [9, 12]. However, weak anomalies are observed at the $T_N$ in the doped samples. The variation of inverse magnetic susceptibility with temperature is shown in inset of figure 4. We found a better description of the paramagnetic susceptibility data by fitting the magnetic susceptibility to the modified Curie – Weiss (CW) law, given by $\chi = \chi_0 + C/T-\theta_{CW}$, where $\chi_0$, C and $\theta_{CW}$ are the temperature independent part of the magnetic susceptibility, Curie constant and Curie-Weiss temperature, respectively. The paramagnetic susceptibility follows the Curie – Weiss law in the range 185 to 300 K for samples x = 0.1 - 0.4. For x = 0.5 sample, a deviation from this behavior occurs at above 280 K as shown in inset of figure 4. This curvature is because of the $T_N$ (~ 350K, estimated from reference [ 25] ) of the of the orthorhombic phase. So for $YMn_{0.5}Fe_{0.5}O_3$ the fitted range has been reduced to 185-250K. All the studied samples follow Curie – Weiss behavior with negative values of Curie temperature indicating the antiferromagnetic interactions. The values of $\chi_0$, Curie constant and the Curie – Weiss temperature ($\theta_{CW}$) obtained from this fit and are summarized in table 1. From the Curie constant values, we have calculated the effective paramagnetic moment ($\mu_{eff}$) by using eq., $C = \frac{N\mu_{eff}^2\mu_B^2}{3k_B}$. The values of $\theta_{CW}$ and $\mu_{eff}$, obtained for $YMnO_3$ are -421K and 4.98$\mu_B$, respectively. The value of $\theta_{CW}$ decreases for ($0.0 \le x \le 0.2$) and then it start increasing for higher doped samples ($0.3 \le x \le 0.5$). The values of $\mu_{eff}$ for x = 0.1, 0.2, 0.3, 0.4 and 0.5 are 4.45$\mu_B$, 4.43$\mu_B$, 4.87$\mu_B$, 5.30$\mu_B$, and 5.23$\mu_B$, respectively. The expected effective moment values were calculated by assuming both Mn and Fe are in trivalent state. Theoretically, $\mu_{eff}$ is calculated as, $\mu_{eff}^{cal} = \sqrt{x\mu_{eff}^2(Fe^{3+}) + (1-x)\mu_{eff}^2(Mn^{3+})}$ where, $\mu_{eff}$ for $Fe^{3+}$ (S= 5/2) and for $Mn^{3+}$ (S=2) are 5.9 $\mu_B$ and 4.89 $\mu_B$ respectively. The expected values of $\mu_{eff}$ for x = 0.1, 0.2, 0.3, 0.4, and 0.5 are 5.01 $\mu_B$, 5.12 $\mu_B$, 5.23 $\mu_B$, 5.33 $\mu_B$, and 5.43 $\mu_B$, respectively. The experimentally obtained magnetic moments are close to the expected values for x>0.2.

**Mössbauer spectrometry**

Mössbauer study was carried out to confirm the oxidation state of iron (Fe) in these doped samples. Figure 5 shows the Mössbauer spectrum for $YMn_{0.8}Fe_{0.2}O_3$ at room temperature. In earlier Mössbauer study of $YMn_{0.9}Fe_{0.1}O_3$, three different sites have been observed for Fe resulting from different chemical environment around Fe ion [12]. The Mössbauer spectrum of $YMn_{0.8}Fe_{0.2}O_3$ which is isostructural with $YMn_{0.9}Fe_{0.1}O_3$ with different Fe content is similarly fitted with three symmetric doublets. These three doublets correspond to three different chemical environments around Fe which are associated with different number of $Mn^{3+}$ ion near neighbors. Explanation for these three different chemical environments around Fe ion has been given in our earlier paper [12]. For higher Fe doped samples we have obtained a mixed orthorhombic and hexagonal phase. In both these structures the environment around Fe atom is totally different. In orthorhombic phase the Fe ion is in six coordinated state while in hexagonal phase Fe is in 5-fold coordination forming $MnO_5$ trigonal bypiramids. So the Mössbauer spectra are rather complicated for these mixed phases. For x= 0.5 sample, the main phase is the orthorhombic phase with a small contribution of the hexagonal phase. In Mössbauer spectrum of $YMn_{0.5}Fe_{0.5}O_3$ at room temperature sextet is observed because of the magnetic contribution of the orthorhombic phase. The Mössbauer spectrum of $YMn_{0.5}Fe_{0.5}O_3$ is fitted with two doublets (related to hexagonal phase) and two sextets (corresponding to the magnetic phase of orthorhombic part) as shown in figure 6. The hyperfine parameters i.e. isomer shift ($\delta$), quadrupole splitting ($\Delta E_Q$), line widths ($\Gamma$) and hyperfine field ($H_{hf}$) obtained from the fit are included in Table 2. The isomer shift (δ) values for all the three doublets in $YMn_{0.8}Fe_{0.2}O_3$ are close to 0.30 mm/s at room temperature and can be attributed to the $Fe^{3+}$ (S = 5/2) in fivefold oxygen coordination. The δ values of both the sextets for $YMn_{0.5}Fe_{0.5}O_3$ are greater that 0.30mm/s, and are 0.40mm/s and 0.60mm/s respectively. This could be an indication of octahedrally coordinated $Fe^{3+}$ [35].

**Magnetic structure**

Neutron diffraction patterns for all studied sample have been recorded at selected temperatures between 6 and 300 K. Figure 7 shows a section of the diffraction data at 300K and 6K for $YMn_{0.8}Fe_{0.2}O_3$. This is a representative of the samples with $x \leq 0.2$. The refinement of room temperature diffraction data has been carried out in $P6_3cm$ space group. Below 65K superlattice reflections (100) (101) and enhancement in the intensity of fundamental Bragg reflection (102) is observed indicating the antiferromagnetic nature of the samples. The magnetic structure has been refined by representation analysis using Sarah program [36]. According to the representation theory there are six different representations (Γ1 – Γ6) which are compatible with space group $P6_3cm$ with propagation vector, $\vec{k} = 0$ [37, 9]. The (100) Bragg peak is a pure magnetic peak in $YMnO_3$ and is only present in Γ3 representation (completely absent for Γ4 representation), while another magnetic peak (101) peak is seen in Γ4 representations. We find the magnetic structure of $YMnO_3$ at 6 K is best described by Γ3 with 26% mixing of the Γ4 representation, with moment (M) = 3.24 $\mu_B$ [12]. In earlier studies, authors have interpreted that the magnetic structure of $YMnO_3$ could be explained by either Γ1 or Γ3 [9, 10]. But according to a theoretical study the magnetic ground state of $YMnO_3$ should be Γ3 not Γ1 [38]. However, recent polarization neutron diffraction studies by Brown and Chatterji [39] and J-G Park et al.[ 11 ] found that the magnetic structure of $YMnO_3$ is better explained by mixed representations. With Fe doping at Mn site, a change in ratio of the intensities of the magnetic reflections is observed. A significant reduction in the intensity of (100) and an enhancement in the intensity of (101) magnetic peaks is seen in 20% Fe doped samples as shown in inset of figure 7. With Fe doping the amount of Γ4 IR increases and the mixing ratio changes to 92% for $YMn_{0.8}Fe_{0.2}O_3$ at 6K, with moment 2.35 $\mu_B$. The reorientation of the spins in Fe doped samples (on changing from Γ3 to Γ4 IR) may be ascribed to the different magnetic anisotropy of the $Fe^{3+}$ and $Mn^{3+}$ ions [25]. The presence of electron in $d_z^2$ orbital in Fe doped $YMnO_3$ influences the anisotropy of the system leading to reorientation of the spins. The thermal variation of the refined Mn/Fe magnetic moments for $YMn_{0.8}Fe_{0.2}O_3$ is shown in the inset of figure 7. We calculate the thermal variations of the moment by applying molecular field model. Using the Brillouin description for reduced magnetization, $m_{Mn} = m_{sat}(T)B_2(x)$, where $x = \frac{mN\lambda(gS\mu_B)^2}{k_B T}$, $m_{sat}$ = 2.35 $\mu_B$, and molecular field constant $\lambda = \frac{3k_B T_c}{g^2 S(S+1)\mu_B^2}$ [40], we obtained a good fit to the experimental data, as shown in inset of figure 7. For these doped sample, we obtained $\lambda$ = 10.2 T/$\mu_B$ by taking $T_N$ = 60 K and S = 2.109. Here we are using $T_N$ as a parameter and obtained a good fit by taking $T_N$ = 60K .We do not observe reduction in $T_N$, because of the absence of neutron diffraction data at smaller intervals of temperature, although small reduction in $T_N$ has been observed in earlier studies [28]. The angle (φ) changes from 11.8° for $YMnO_3$ to 55° for $YMn_{0.8}Fe_{0.2}O_3$ (moment on Mn is inclined at 55° to the a axis while for $YMnO_3$ it is inclined at 11.8° to the a axis) at 6K. Enhancement in angle (φ) has been observed as a function of Fe concentration as shown in figure 8. The parent sample ($YMnO_3$) and Fe doped samples shows different behavior of angle (φ) as a function of temperature, as shown in the inset of figure 8. In case of $YMnO_3$, φ angle increases with increase in temperature whereas for Fe-doped samples φ angle decreases with increase in temperature. Similar, spin reorientation has also been observed in $HoMnO_3$. But the spin reorientation in $HoMnO_3$ is very abrupt and occurs at a specific temperature unlike in our case where the transition is continuous [41]. The $Mn^{3+}$ spins in $HoMnO_3$ reorient sharply by $90^{\circ}$ at the spin reorientation temperature. The spin structures for $YMnO_3$ and $YMn_{0.8}Fe_{0.2}O_3$ at 6 K, are shown in figure 9. For both the samples, the spins are coupled ferromagnetically along c-axis but in case of $YMn_{0.8}Fe_{0.2}O_3$ the spins are more tilted away from the a axis. In contrast, in the case of $YbMnO_3$, enhancement in transition temperature ($T_N$) has been observed with Fe doping at Mn site [23] and the magnetic structure, though not analyzed by the authors but we infer, remain the same with Fe doping, albeit with an enhancement in the moment values at Mn site.

In higher Fe doped samples i.e. for $0.3 \leq x \leq 0.5$, a two phase refinement of neutron diffraction data has been carried out taking into account orthorhombic (*Pnma*) and hexagonal (*P6₃cm*) phases. Figure 10 shows a section of the diffraction data at 300K and 6K for $YMn_{0.7}Fe_{0.3}O_3$. On lowering temperature below 300K two superlattice reflections (110) and (011) appear in the diffraction pattern. These two reflections are identified with the magnetic peaks corresponding to orthorhombic phase and are similar to the end member of this series i.e $YFeO_3$. To represent the magnetic structure of the orthorhombic phase Bertaut's notation in the *Pnma* setting is adopted [37, 42]. The spin structure in this temperature range is explained by Γ1 ($G_xC_yA_z$) in *Pnma* setting. The $C_y$ and $A_z$ components are found to be very small. In addition, for x = 0.3 and 0.4 samples, superlattice reflections (100) and (101) also appears below 65K which are characteristics of the magnetic phase of the $YMnO_3$ compound. For $YMn_{0.7}Fe_{0.3}O_3$ sample the magnetic structure in hexagonal phase is explained by taking 71.7% of Γ3 and 28.3% of Γ4. In $YMn_{0.6}Fe_{0.4}O_3$, we are not able to determine the amount of Γ3 and Γ4 IRs accurately, due to small amount of hexagonal phase. The magnetic ground state of $YMn_{0.7}Fe_{0.3}O_3$ and $YMn_{0.6}Fe_{0.4}O_3$ is explained by taking $\Gamma 1$ ($G_xC_yA_z$) representation of *Pnma* setting and $\Gamma 3+\Gamma 4$ representations of *$P6_3cm$* symmetry. In $YMn_{0.5}Fe_{0.5}O_3$ the major phase is the orthorhombic phase with a very small contribution of the hexagonal phase. So the magnetic contribution for this sample is mainly from the orthorhombic phase. Figure 11 shows a section of the neutron diffraction data at 300K and 6K for x=0.5 which is representative of these samples. The spin structure in this temperature range is explained by $\Gamma 1$ ($G_xC_yA_z$) in *Pnma* setting. The moment in the orthorhombic phase for $YMn_{0.5}Fe_{0.5}O_3$ is found to be 2.80 $\mu_B$ at 6K, which is lower than the expected value. Similar reduction in ordered moment value has been seen in $YFe_{0.6}Mn_{0.4}O_3$ [25]. The magnetic structure of $YFeO_3$ is described by taking $\Gamma 4$ ($A_xF_yG_z$) (IR) where $G_z$ represents the antiferromagnetic arrangement of $Fe^{3+}$ spins along the z-axis and $F_y$ represents the ferromagnetic arrangement of spins along the y-axis due to the canting of $G_z$ spins [33,44]. Doping with Mn is found to change the antiferromagnetic easy axis from z- axis $\Gamma 4$ ($A_xF_yG_z$) to x- axis $\Gamma 1$ ($G_xC_yA_z$) in $YFe_{1-x}Mn_xO_3$ ($0.10 \leq x \leq 0.45$) and has been ascribed to the different magnetic anisotropy of $Fe^{3+}$ and $Mn^{3+}$ ions [25]. Mn doping also leads to enhancement in $T_{SR}$ (spin reorientation temperature) and reduction in $T_N$. For $YFe_{0.55}Mn_{0.45}O_3$ sample, $T_{SR}$ is 330K, which means the antiferromagnetic easy axis changes from z- axis $\Gamma 4$ ($A_xF_yG_z$) to x- axis $\Gamma 1$ ($G_xC_yA_z$) at this temperature. Extrapolating $T_{SR}$ to higher compositions of Mn indicates that $T_{SR}$ would be above 300K for x=0.5 sample. This explains the absence of $T_{SR}$ in the studied samples and the magnetic structure at 300K is already $\Gamma 1$.

**Dielectric properties**

The dielectric constant (ε′) as a function of temperature (behavior is same for all frequencies, for clarity only 10 kHz data shown) is shown in figure 12. A distinct anomaly is observed in dielectric constant and tanδ for $YMnO_3$ near Néel temperature ($T_N$) as shown in figure 12(a). This behavior is similar to that reported earlier for the same compound [4, 6, 45]. For a detailed investigation of magnetoelectric effect in Fe doped $YMnO_3$, we chose two compositions, one with the hexagonal structure (x = 0.2) and one with the orthorhombic phase (with a small contribution of the hexagonal phase, x = 0.5) as shown in figure 12(b) and 12(c). With Fe doping the anomaly is observed to be diminished as shown in figure 12(b). Below $T_N$, however, no further discontinuity in ε′ is observed in the x=0.2 doped sample, in which we observe a spin reorientation behavior as a function of temperature. This is in contrast to dielectric measurement of single crystal of $HoMnO_3$, a peak has been observed at the spin reorientation temperature [21, 46], where the $Mn^{3+}$ spin reorient sharply by $90^{\circ}$. No such anomaly is observed in the case of $YMnO_3$ and Fe doped samples because in these the orientation of the spin changes continuously with temperature. This is in agreement with the progressive suppression of the spin reorientation behavior observed in the case of $Ho_{1-x}Y_xMnO_3$ samples [47]. The coupling of ferroelectric and magnetic order in a ferroelectromagnets, in the absence of an external applied field occurs by the electron-phonon interactions [48, 49]. Here in case of an antiferromagnetic $YMnO_3$ in the absence of an external electric and magnetic field, isostructural phase transition at antiferromagnetic transition temperature ($T_N$) and atomic displacements of atoms below $T_N$ are likely to be responsible for observed magnetoelectric effect [4,49]. The displacement of Mn from its x~1/3 position at $T_N$ and below has been observed in previously reported neutron diffraction experiments [50]. However, with our experimental resolution we do not find the displacement of the Mn atom. The suppression of dielectric anomaly in Fe doped samples may be correlated with the change in coupling between spin correlation and electric polarization. In earlier studies of Fe doping at Mn-site of $YbMnO_3$, it has been observed that increase in Fe content leads to weakening of ferroelectricity in the system [23]. For the ferroelectric distortion to occur, the d orbitals in the direction of electric polarization must be empty. The ferroelectric distortion in hexagonal manganites is induced by the hybridization between unoccupied $d_z^2$ orbital of Mn and $p_z$ orbital of oxygen atom. The weakening of ferrolectricity in Fe doped $YbMnO_3$ has been attributed to the presence of partially filled $dz^2$ orbital which lowers the degree of hybridization between $d_z^2$ orbital of Mn/Fe and $p_z$ orbital of oxygen atom. In $YMn_{0.5}Fe_{0.5}O_3$, the main phase is the orthorhombic phase with a very small contribution from the hexagonal phase (~ 7%), so the dielectric anomaly in this compound might be related with the remnant hexagonal phase of the $YMnO_3$. As shown in the inset of figure 12(a), two semicircle arcs can be seen in the impedance plane. Similar kind of two semicircular arcs has been seen in case of $HoMnO_3$ ceramic samples [51]. These two arcs in different frequency regions correspond to contribution from sample electrodes interface as well as from sample. The arc in the lower frequency region (away from the origin) corresponds to the dielectric response from electrodes, whereas the dielectric response from grain and grain boundary is represented by the arc in high frequency region (towards the origin). The complex impedance can be defined as

$$Z^* = Z' + jZ'' = \frac{R}{1+(\omega CR)^2} + j\frac{\omega CR^2}{1+(\omega CR)^2},$$

Where $Z^*$, R and C are the complex impedance, resistance and capacitance of the studied samples. Two semicircular arcs are also observed in Fe doped samples as shown in inset of figure 12(b) and figure 12(c).

The anomaly in dielectric constant at Néel temperature ($T_N$) for x=0, 0.2 &0.5 ($YMn_{1-x}Fe_xO_3$) samples is evidenced by a decrease in ε below the $T_N$. In order to analyze this dielectric anomaly at $T_N$, we have obtained the value dielectric constant at 0K, ε´(0) using a exponential function similar to that done before [ 52]. Then we subtract off the lattice contribution from the observed dielectric constant below $T_N$ and this difference $\Delta\varepsilon$ is shown in inset of figure 13 for these three samples. We find a different behavior of dielectric difference as a function of temperature for parent and Fe doped samples. With Fe doping the difference between lattice contribution and the observed dielectric constant decreases. This behavior of dielectric anomaly below $T_N$ may be attributed to the different spin structure for parent and Fe doped sample. To understand the magnetoelctric effect of ferroelectric antiferromagnet, we have considered the free energy expansion in the Landau theory [52].

$$F = F_0 + \frac{a_1}{2}L^2 + \frac{a_2}{2}L^4 + \frac{b_1}{2}P^2 + \frac{c_1}{2}P^2L^2 + \frac{c_2}{2}P^2H^2 - EP$$

where E, H, P and L are the electric field, magnetic field, polarization and antiferromagnetic vector, respectively. The minimum of the free energy defines the equilibrium state of the system. By considering the equilibrium condition, we find

$$\frac{\partial F}{\partial P} = 0, \qquad \frac{\partial F}{\partial L} = 0$$

$$\frac{\partial F}{\partial P} = b_1P + c_1PL^2 + c_2PH^2 - E = 0.$$

Since, dielectric constant of a medium can be defined as, $\varepsilon = 1 + \frac{P}{\varepsilon_0 E}$ where $\varepsilon_0$ is the permittivity of the free space. Using above equation in the absence of any external magnetic field (H=0), for small values of L the dielectric constant can be written as, $\varepsilon \approx 1 + \frac{1}{b_1} - \frac{c_1L^2}{b_1^2}$. We can rewrite this equation as $\varepsilon = a + bL^2$. Therefore, the temperature dependence of dielectric permittivity near $T_N$ is expected to be proportional to the $L^2(T)$. This antiferromagnetic vector (L) is equal to the magnetic moment (M) below $T_N$. In figure 13 we show a plot of $\Delta\varepsilon$ as a function of $M^2$. A clear linear region is observed for low M values demonstrating a coupling of the two order parameters in the absence of field. Earlier, a similar suppression of dielectric behavior in the magnetically ordered state and a linear dependence of $\Delta\varepsilon$ versus $M^2$ had been observed in $BaMnF_4$ [53] and attributed to presence of weak ferromagnetic character in the sample. Similar behavior has also been reported in the case of non-centrosymmetric ferromagnet $BiMnO_3$ [54]. In contrast to the previous observations, the present results bring out a correlation between $\varepsilon$ and M in the case of a frustrated antiferromagnet in the absence of a magnetic field and ferromagnetic behavior.

## Conclusion

We have synthesized polycrystalline samples of $YMn_{1-x}Fe_xO_3$ (x = 0.1, 0.2, 0.3, 0.4, 0.5) and studied their structural, magnetic and dielectric properties. A single hexagonal phase has been observed for $x \leq 0.2$ and for $0.3 \leq x \leq 0.5$ a mixed hexagonal (characteristic to $YMnO_3$) and orthorhombic phase (characteristic to $YFeO_3$) has been obtained. The magnetic structure in case of isostructural compounds i. e. for $x \leq 0.2$ is explained by taking linear combination of $\Gamma3$ and $\Gamma4$ IR but with different mixing ratios of these two representations. With Fe doping at Mn site spin reorientation has been observed, the angle ($\varphi$) changes from $11.8^\circ$ for $YMnO_3$ to $55^\circ$ for $YMn_{0.8}Fe_{0.2}O_3$ at 6K. In orthorhombic phase the magnetic structure is explained by $\Gamma1$ ($G_xC_yA_z$) IR of *Pnma* setting In $YMnO_3$ suppression of dielectric constant $\varepsilon'$ is observed below $T_N$ indicative of magnetoelctric coupling. This anomalous behavior reduces in Fe doped samples, The dielectric constant is found to be correlated with the magnetic moment (M) obtained from neutron diffraction experiments and follows a $M^2$ behavior close to $T_N$ in agreement with Landau theory.

**Figure Captions:**

**Fig 1:** The variation of volume fraction of hexagonal and orthorhombic phases of $YMn_{1-x}Fe_xO_3$ with the Fe content x. Hexagonal and orthorhombic phases are indicated by filled (●) and open (○) symbols, respectively.

**Fig 2:** Temperature variation of the lattice parameters a (○) and c (●). (b) Temperature variation of unit cell volume for $YMn_{0.8}Fe_{0.2}O_3$. Inset shows the tilting and buckling angles at 6 K and 300K.

**Fig 3: (a)** Temperature variation of unit cell volume for $YMn_{0.6}Fe_{0.4}O_3$. The solid line is a fit to the Debye - Grüneisen equation.

**Fig 4:** The zero field-cooled (ZFC) magnetization (M) versus temperature (T) in field of H = 0.1T for $YMn_{1-x}Fe_xO_3$ (x = 0, 0.1, 0.2, 0.3, 0.4, 0.5). Inset shows the inverse of susceptibility as a function of temperature and modified Curie- Weiss fit (solid line).

**Fig 5:** The plot of Mössbauer spectrum of $YMn_{0.8}Fe_{0.2}O_3$ at 300K.

**Fig 6:** The plot of Mössbauer spectrum of $YMn_{0.5}Fe_{0.5}O_3$ at 300K.

**Fig 7:** The observed (○) and calculated (—) neutron diffraction pattern for $YMn_{0.8}Fe_{0.2}O_3$ compound at T = 6 K and 300 K. Lower solid line is the difference between observed and calculated pattern. The first row of tick marks indicates the position of nuclear Bragg peaks and second row indicate the position of magnetic Bragg peaks. Inset (a) shows the variation magnetic

moment as a function of temperature. Inset (b) shows the raw neutron data for x= 0, 0.1 and 0.2 samples at 6K.

**Fig 8:** The variation of tilting angle (φ) as a function of Fe content at 6K. Inset shows the variation of tilting angle (φ) as a function of temperature for $YMnO_3$ and for $YMn_{0.8}Fe_{0.2}O_3$.

**Fig 9:** The magnetic structures for $YMnO_3$ and for $YMn_{0.8}Fe_{0.2}O_3$ at 6 K.

**Fig10:** The observed (○) and calculated (—) neutron diffraction pattern for $YMn_{0.7}Fe_{0.3}O_3$ compound at T = 6 K and 300 K. Lower solid line is the difference between observed and calculated pattern. The first and second row of tick marks indicates the position of nuclear Bragg peaks for hexagonal and orthorhombic phase respectively. The third and fourth rows indicate the position of magnetic Bragg peaks. Inset shows the variation of tilting angle (φ) as a function of temperature for $YMn_{0.7}Fe_{0.3}O_3$.

**Fig11:** The observed (○) and calculated (—) neutron diffraction pattern for $YMn_{0.5}Fe_{0.5}O_3$ compound at T = 6 K and 300 K. Lower solid line is the difference between observed and calculated pattern. The first row of tick marks indicates the position of nuclear Bragg peaks and second row indicate the position of magnetic Bragg peaks. Inset shows the magnetic structure of this compound at 6K.

**Fig 12:** The variation of dielectric constant (ε') with temperature (T) for (a) $YMnO_3$ (b) $YMn_{0.8}Fe_{0.2}O_3$ and (c) $YMn_{0.5}Fe_{0.5}O_3$. Inset shows the complex impedance (Z'') plane plot versus real part of impedance (Z'). Dielectric loss (tan(δ)) versus temperature (T) is shown in the inset of a.

**Fig 13**: The variation of Δε (ε'(T) - ε'(0)) as a function of $M^2(T)$ for (a) YMnO3 and (b) $YMn_{0.8}Fe_{0.2}O_3$. Inset shows the thermal variation of Δε for $YMnO_3$ (x=0), $YMn_{0.8}Fe_{0.2}O_3$ (x=0.2), and for $YMn_{0.5}Fe_{0.5}O_3$ (x=0.5).

**Table Captions**

**Table 1.** Results of Rietveld refinement of neutron diffraction pattern at 6K, transition temperature ($T_N$), parameters obtained from fit to the Curie – Weiss law, and geometrical frustration parameter for $YMn_{1-x}Fe_xO_3$ ($0 \leq x \leq 0.5$).

**Table 2.** The fitted parameters obtained from analysis of the Mössbauer spectra of $YMn_{1-x}Fe_xO_3$ (x = 0.2, & 0.5) at 300K.

**Figure 1**

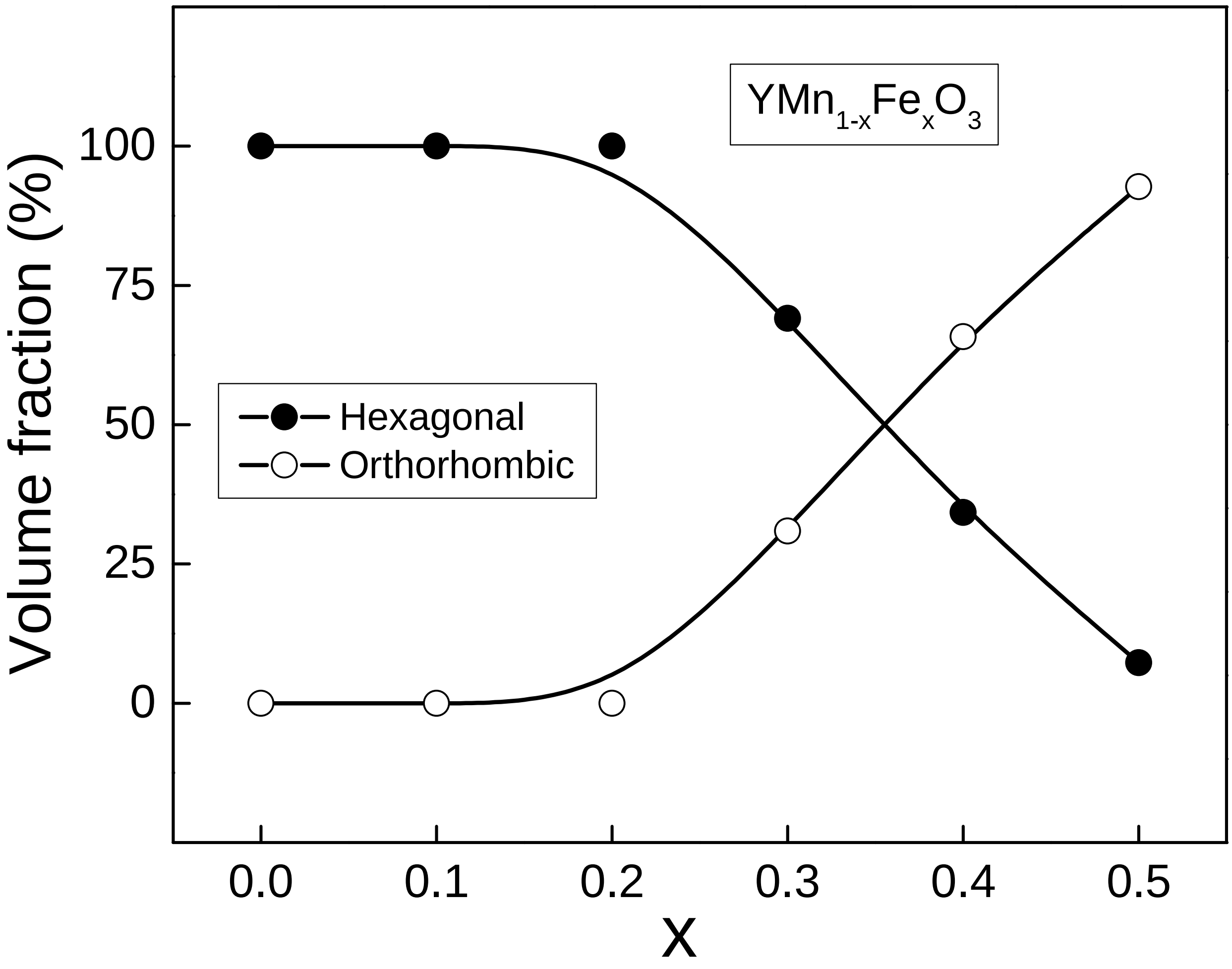

**Figure 2**

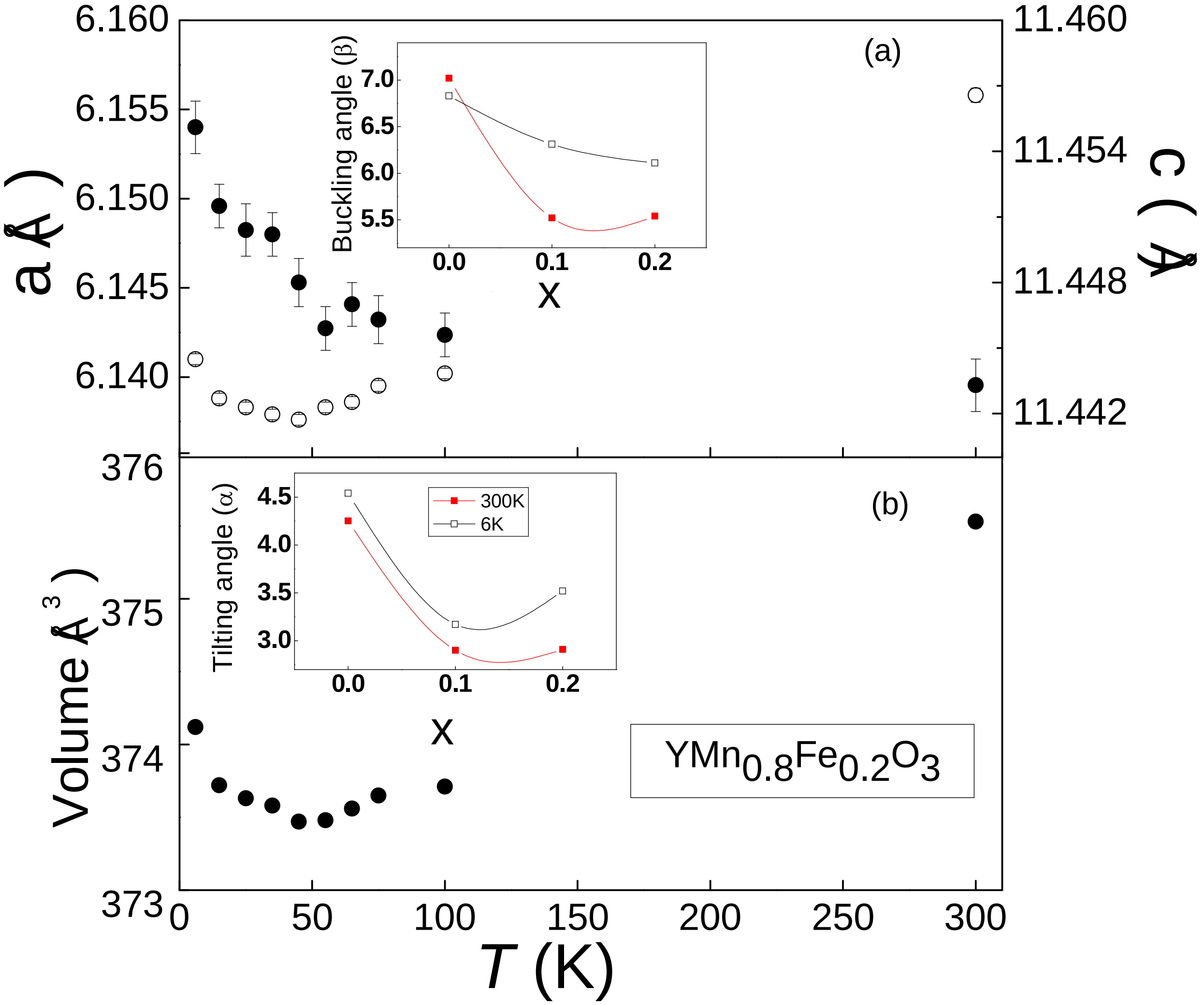

**Figure 3**

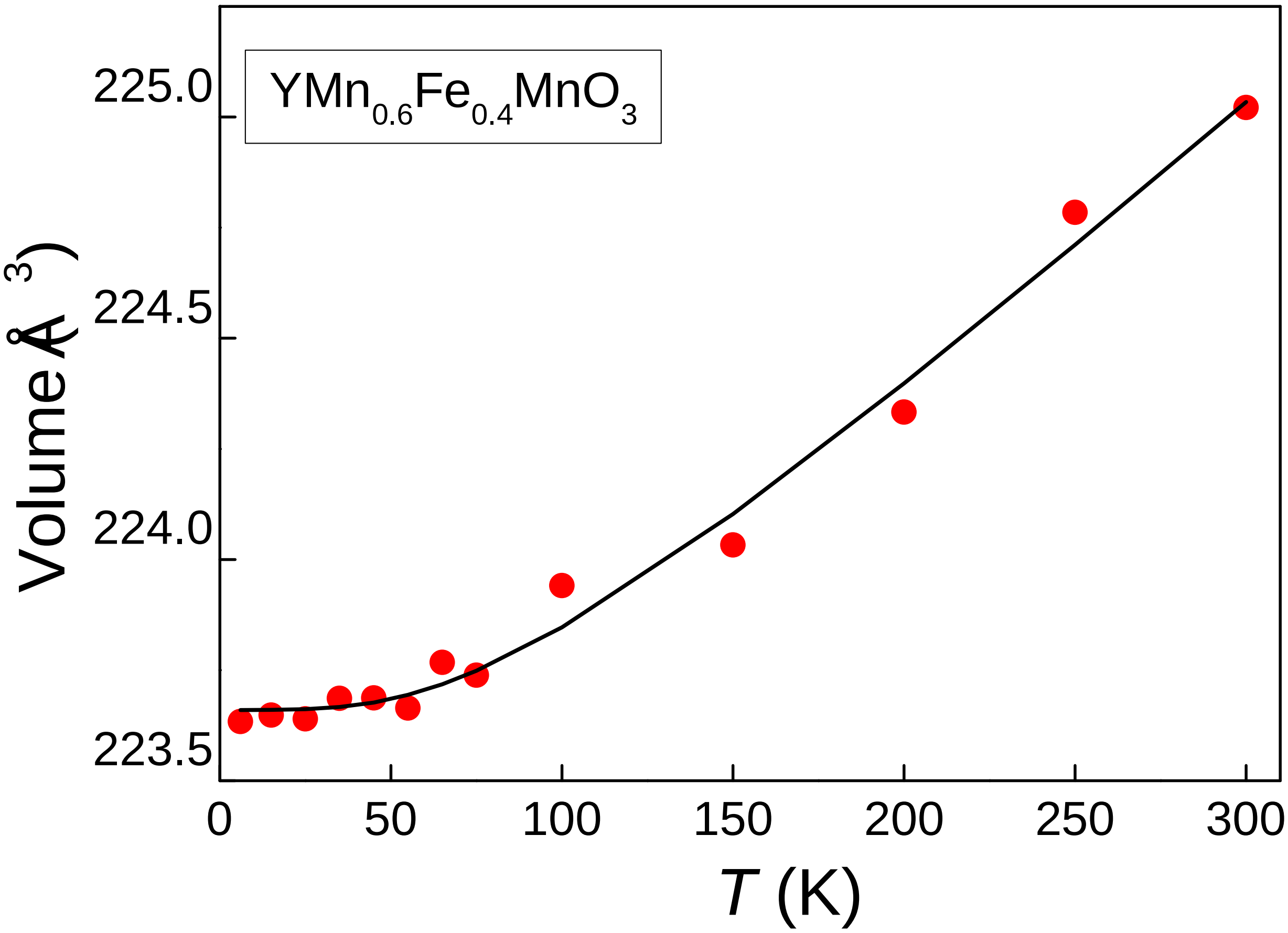

**Figure 4**

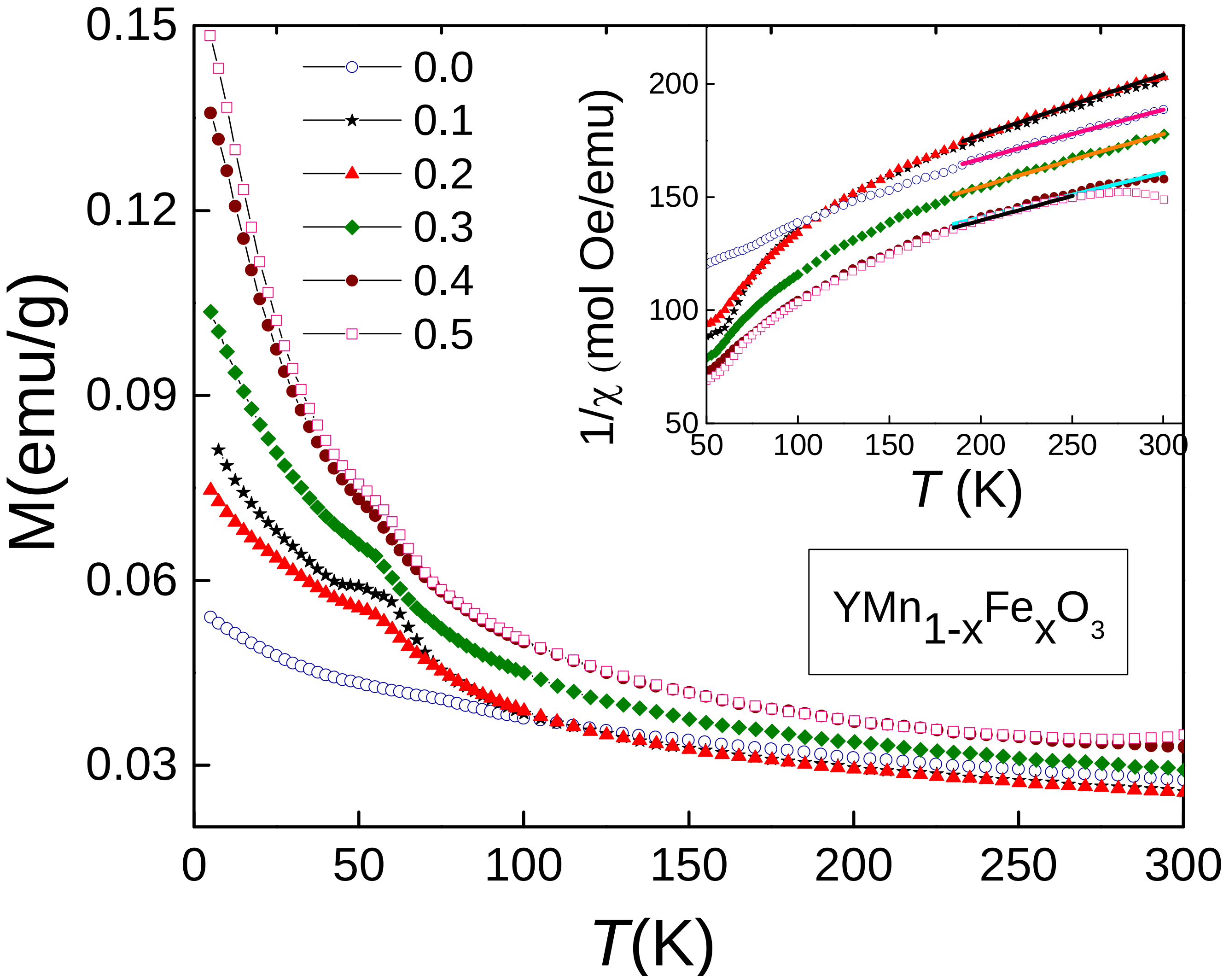

**Figure 5**

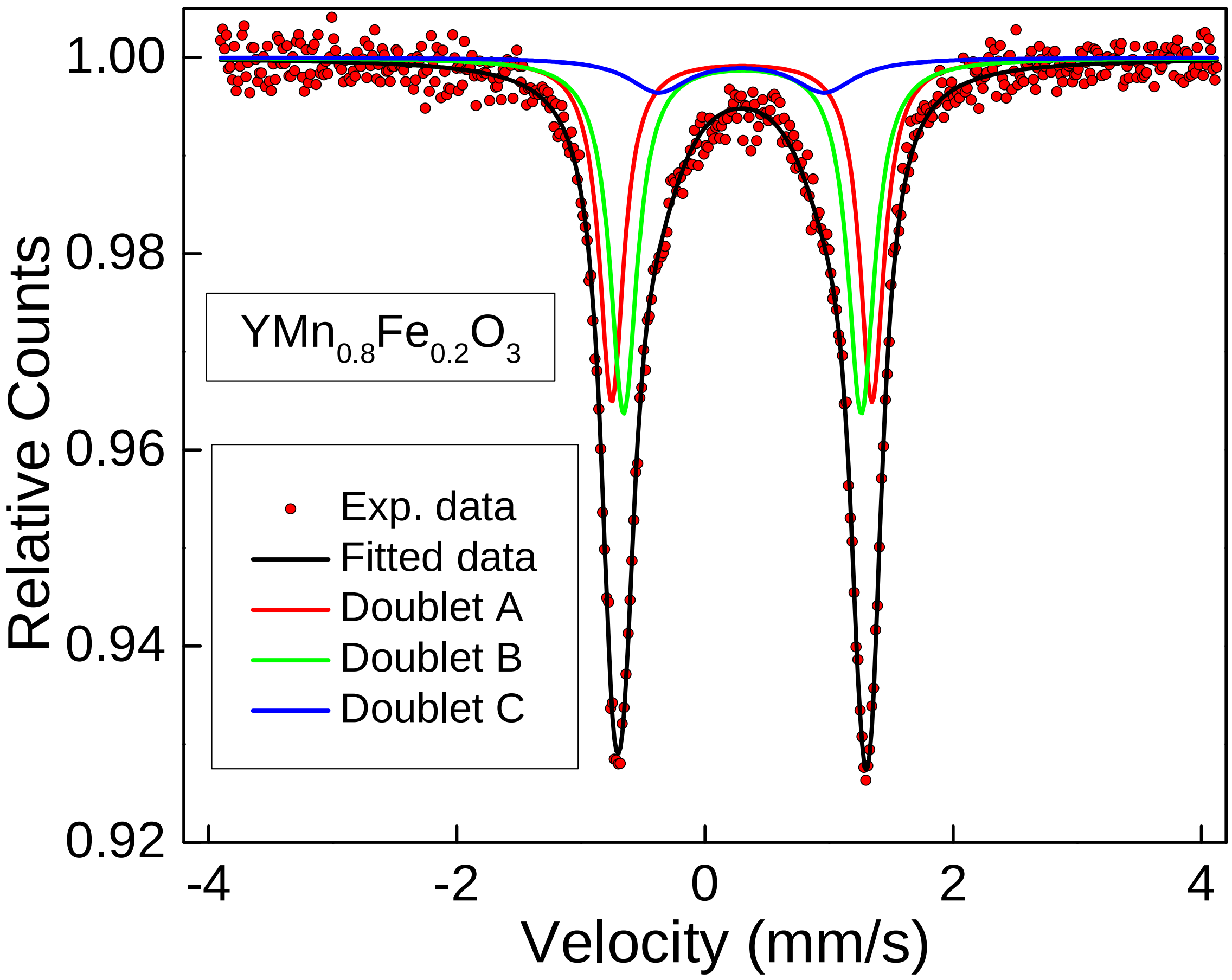

**Figure 6**

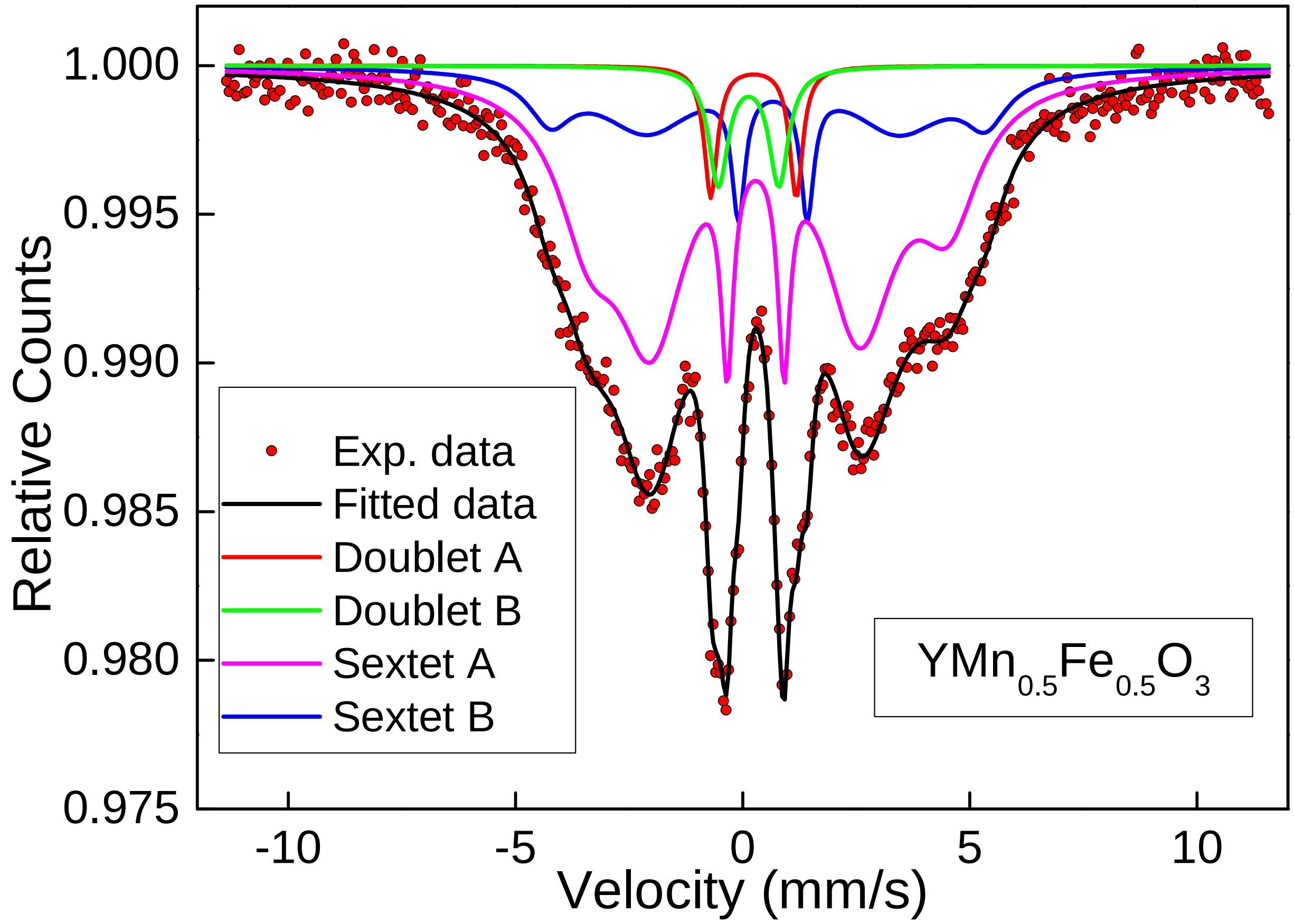

**Figure 7**

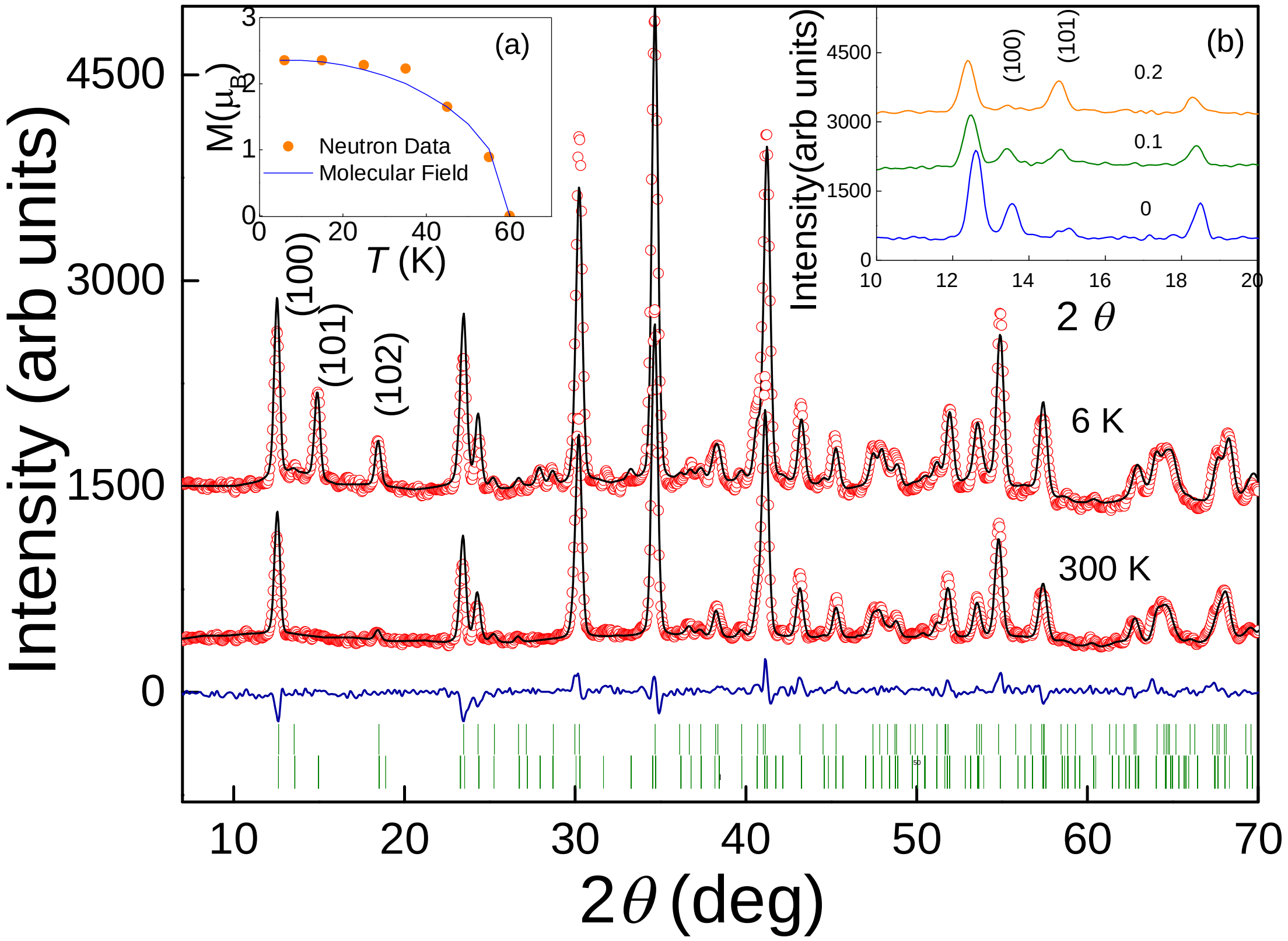

**Figure 8**

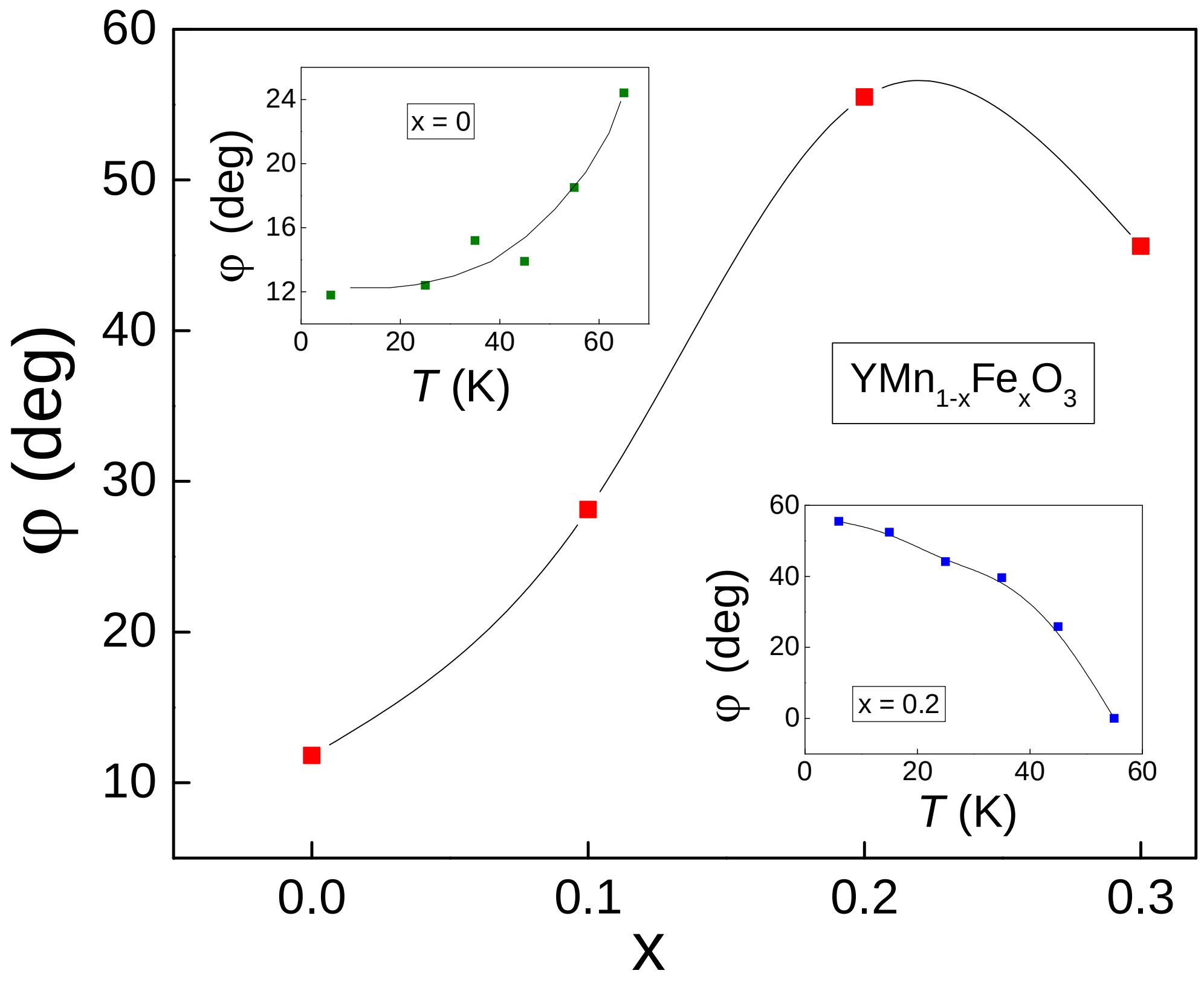

**Figure 9**

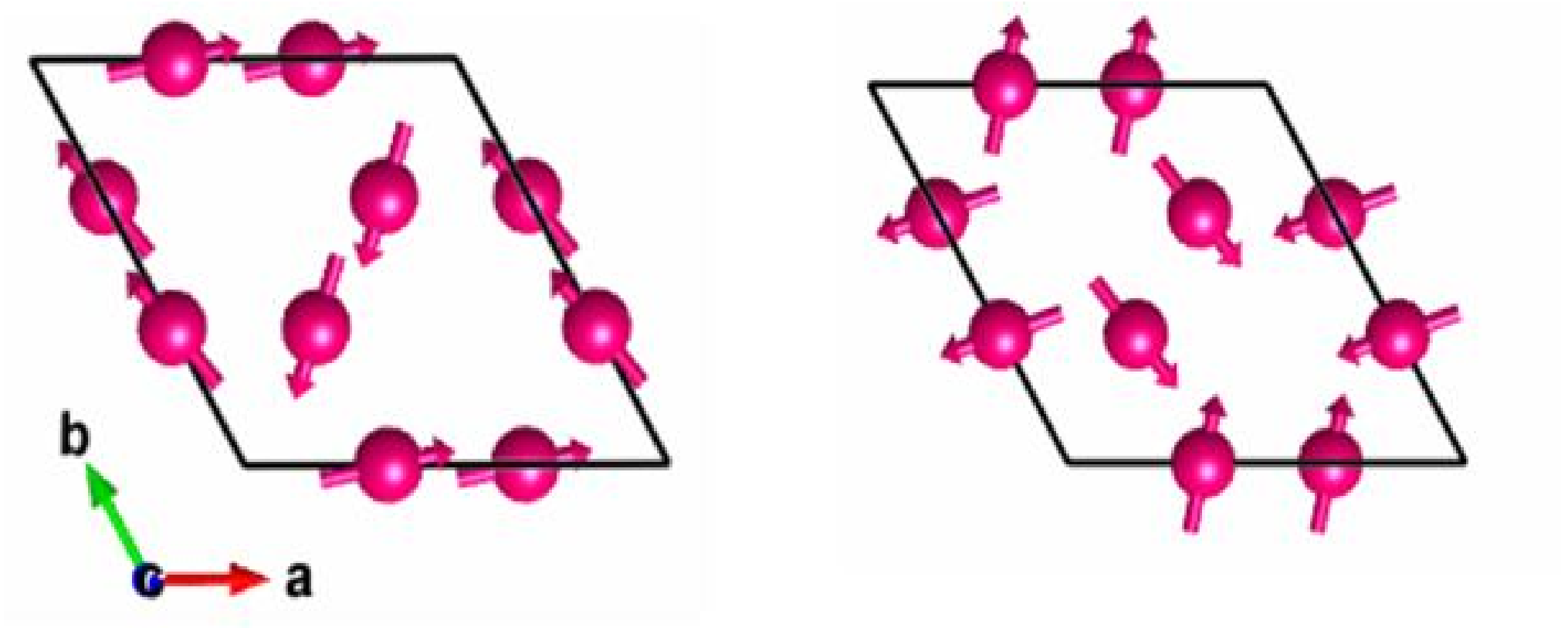

$YMnO_3$ ($\Gamma3$+ 26% of $\Gamma4$) $YMn_{0.8}Fe_{0.2}O_3$ ($\Gamma3$+ 92% of $\Gamma4$)

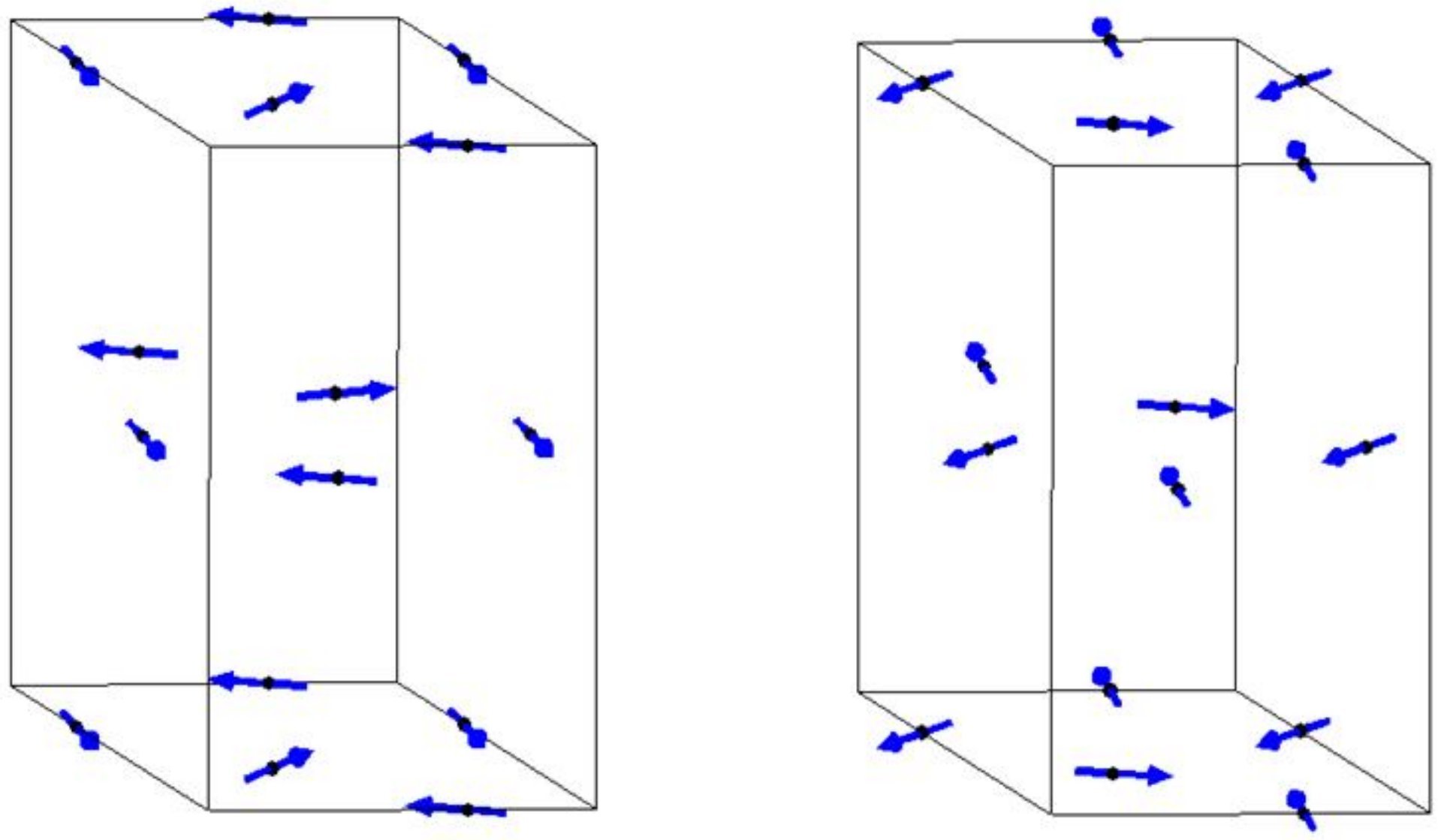

**Figure 10**

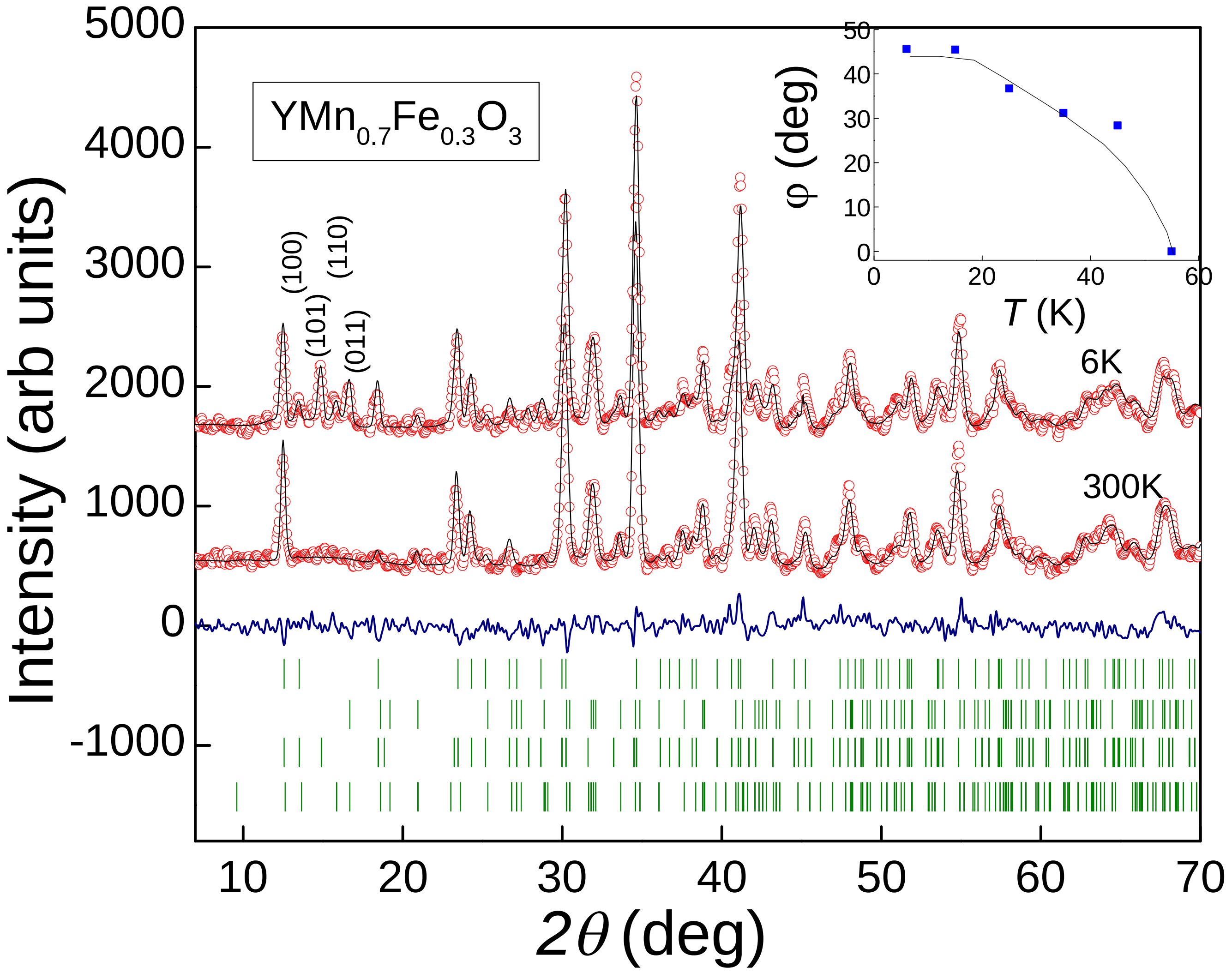

**Figure 11**

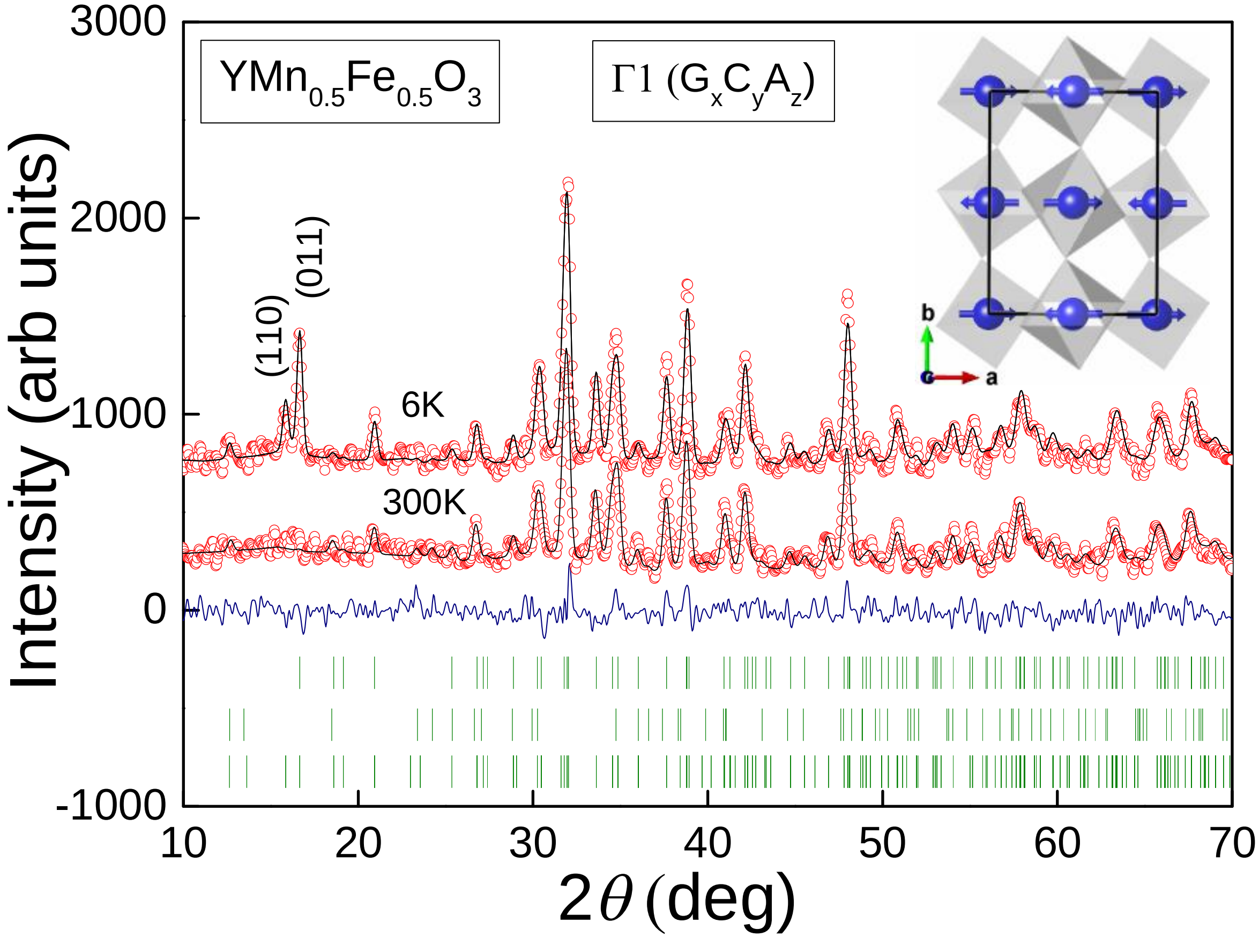

**Figure12**

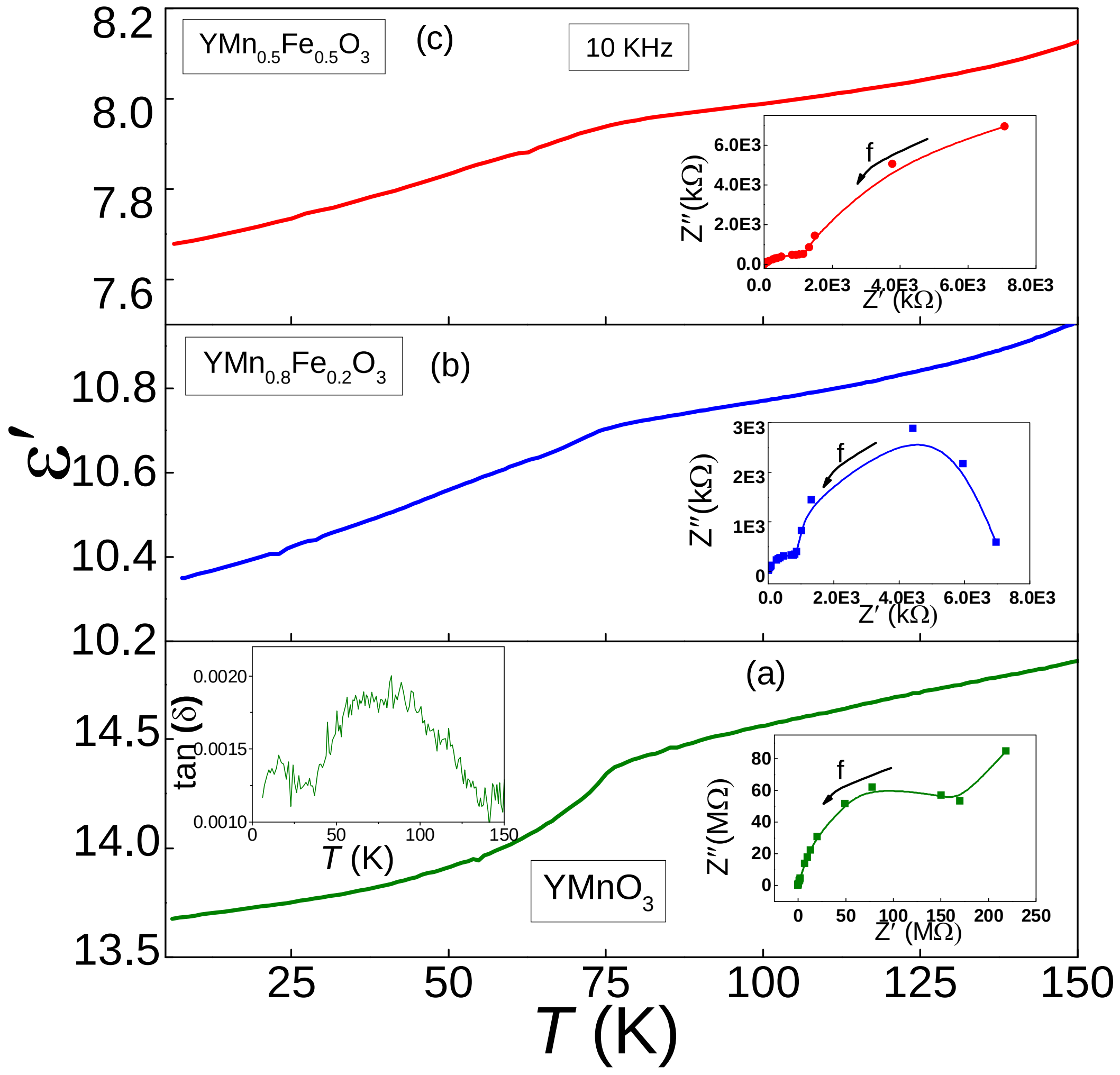

**Figure 13**

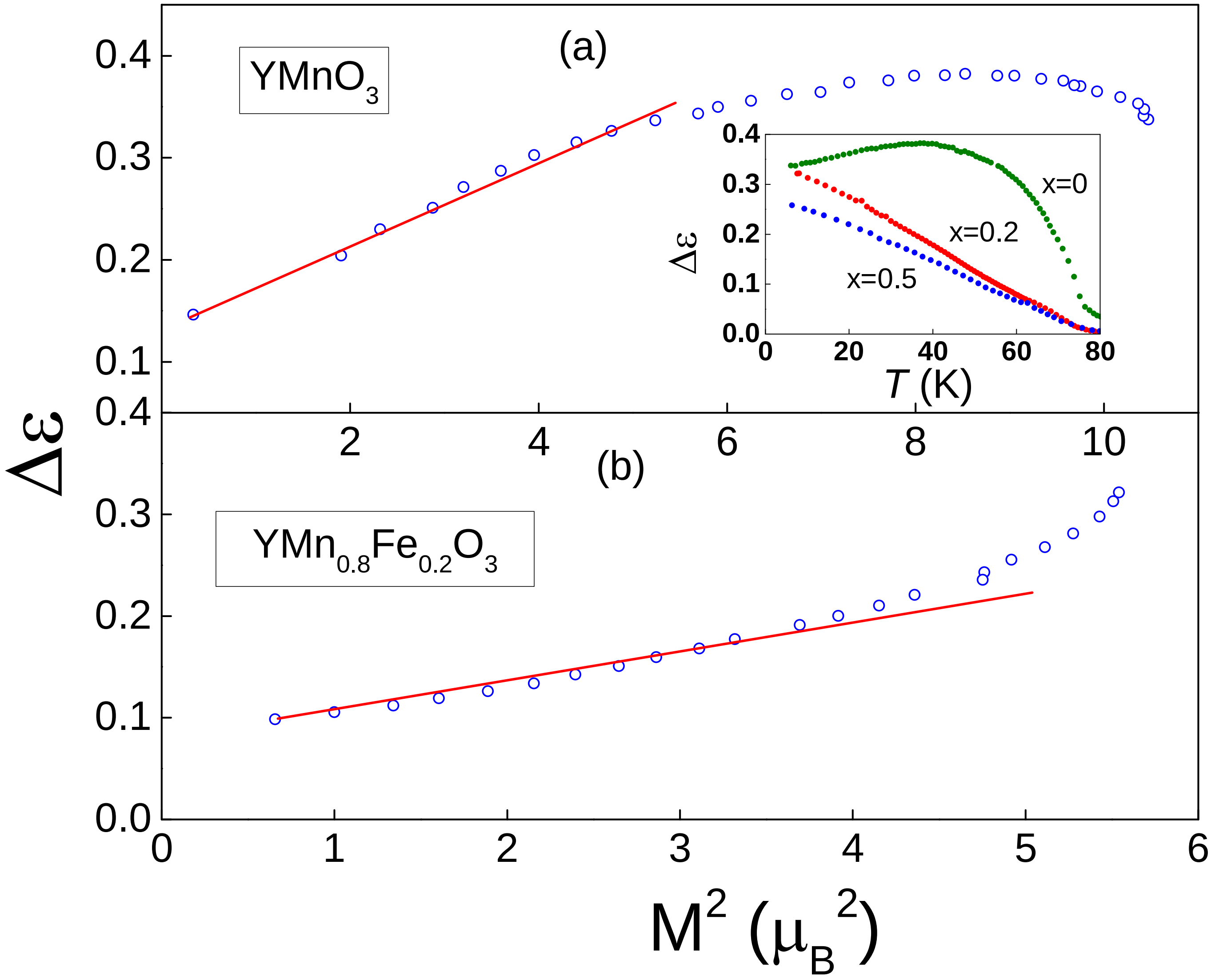

**Table 2.** The Mössbauer parameters for $YMn_{1-x}Fe_xO_3$ (x = 0.2, & 0.5) at 300K.

| **x** | **Iron Sites** | **Isomer shift (*δ*) mm/s** | **Quadrupole splitting ($\Delta E_Q$) mm/s** | **Line width (*Γ*) mm/s** | **Hyperfine Field $H_{hf}$ (Tesla)** | **Area $R_A$ (%)** | **Fitting quality ($\chi^2$)** |
|---|---|---|---|---|---|---|---|
| 0.2 | Doublet A | 0.301±0.002 | 2.101 ±0.032 | 0.206 ±0.033 | -- | 28.2 | |
| | Doublet B | 0.302±0.001 | 1.916 ±0.035 | 0.249 ±0.028 | -- | 52.1 | 1.031 |
| | Doublet C | 0.288±0.009 | 1.291 ±0.046 | 0.449 ±0.042 | -- | 19.7 | |
| 0.5 | Doublet A | 0.238±0.046 | 1.879±0.033 | 0.352±0.05 | -- | 2.2 | |
| | Doublet B | 0.129±0.034 | 1.33±0.032 | 0.520±0.046 | -- | 4.4 | 1.045 |
| | SextetA | 0.601±0.02 | -0.109±0.05 | 0.349±0.05 | 29.71±0.12 | 24.6 | |
| | Sextet B | 0.429±0.012 | 0.289±0.021 | 0.329±0.038 | 24.53±0.16 | 68.8 | |

**Table 1.** Results of Rietveld refinement of neutron diffraction pattern at 6K, transition temperature, Curie – Weiss fit parameters, geometrical frustration parameter for $YMn_{1-x}Fe_xO_3$ ($0.1 \leq x \leq 0.5$).

| | **$YMnO_3$** (Ref 12) | **$YMn_{0.9}Fe_{0.1}O_3$** (Ref 12) | **$YMn_{0.8}Fe_{0.2}O_3$** | **$YMn_{0.7}Fe_{0.3}O_3$** | | **$YMn_{0.6}Fe_{0.4}O_3$** | | **$YMn_{0.5}Fe_5O_3$** | |
|---|---|---|---|---|---|---|---|---|---|
| | | | | *$P6_3cm$* | *Pnma* | *$P6_3cm$* | *Pnma* | *$P6_3cm$* | *Pnma* |
| a (Å) | 6.1212(4) | 6.1359(4) | 6.1410(4) | 6.1395(4) | 5.691(2) | 6.1323(11) | 5.6797(11) | 6.1550(16) | 5.6805(8) |
| b (Å) | 6.1212(4) | 6.1359(4) | 6.1410(4) | 6.1395(4) | 7.4909(20) | 6.1323(11) | 7.4860(14) | 6.1550(16) | 7.50146(11) |
| c (Å) | 11.4002(9) | 11.4289(9) | 11.4551(12) | 11.4476(15) | 5.2646(15) | 11.425(3) | 5.2597(8) | 11.3540(21) | 5.2666(9) |
| V ($Å^3$) | 369.93 | 372.64 | 374.12 | 373.68 | 224.43 | 372.09 | 223.63 | 372.51 | 224.42 |
| Mn-O1 (Å) | 1.90 (2) | 1.88 (3) | 1.81 (6) | 1.97(6) | - | 2.37(6) | - | | |
| Mn-O2 (Å) | 1.86 (2) | 1.82 (3) | 1.90(5) | 1.71(9) | - | 1.356(19) | - | | |
| Mn-O3 (Å) | 2.082 (3) | 2.08 (2) | 2.10(5) | 2.18(6) | - | 2.477(16) | - | | |
| Mn-O4 (Å) | 2.039 (3) | 2.043 (16) | 2.04(2) | 2.02(5) | - | 1.954(17) | - | | |
| Mn-O3-Mn (°) | 119.24 (12) | 119 (2) | 119.1(5) | 119.9(7) | - | 119.49(12) | - | | |
| Mn-O4-Mn (°) | 118.51 (11) | 119.1(6) | 119.2(6) | 116.4(7) | - | 111.70(20) | - | | |
| Mn-$O_1$ (Å) | - | - | - | | 1.97(8) | - | 1.98(6) | | 1.989(16) |
| Mn-$O_{21}$ (Å) | - | - | - | | 2.13(7) | - | 2.11(5) | | 2.107(15) |
| Mn-$O_{22}$ (Å) | - | - | - | - | 1.96(8) | - | 1.97(7) | | 1.966(20) |
| $T_N$ (K) | 75 | 60 | 60 | 55 | 250 | 55 | 250 | 250 | |
| θ (K) | - 421 | -334 | -328 | -343 | | -374 | | -351 | |
| $\mu_{eff}$ ($\mu_B$) | 4.98 | 4.45 | 5.12 | 4.87 | | 5.30 | | 5.21 | |
| $\chi_0$ (emu/mol Oe) | 0.001 | 0.001 | 0.001 | 0.001 | | 0.001 | | 0.001 | |
| f= ($\theta/T_N$) | 5.6 | 5. 6 | 5.46 | - | | - | | | |